# AN INFORMATION SYSTEM FOR INTEGRATED LAND AND WATER RESOURCES MANAGEMENT IN THE KARA RIVER BASIN (TOGO AND BENIN)

Hèou Maléki BADJANA[1,2,3], Franziska ZANDER[3], Sven KRALISCH[3,4],
Jörg HELMSCHROT[4,5], Wolfgang-Albert FLÜGEL[3,4]

[1]West African Science Service Center on Climate Change and Adapted Land Use (WASCAL, www.wascal.org), Graduate Research Program – Climate Change and Water resources, University of Abomey-Calavi, Cotonou, Benin
[2]Laboratory of Botany and Plant Ecology, Faculty of Sciences; University of Lome, Lome, Togo
[3]Department of Geoinformatics, Hydrology and Modelling, Friedrich-Schiller University of Jena, Jena, Germany
[4]Southern African Science Service Center for Climate Change and Adaptive Land Management (SASSCAL, www.sasscal.org)
[5]Biocentre Klein Flottbek and Botanical Garden, University of Hamburg, Hamburg, Germany

## ABSTRACT

*A prerequisite for integrated land and water resources management (ILWRM) is a holistic river basin assessment. The latter requires information and data from different scientific disciplines but also appropriate data management systems to store and manage historical and real time data, set up protocols that facilitate data and information access and sharing among different stakeholders, and triggering further collaboration among different institutions in support of watershed-based assessment, management and planning. In West Africa in general and especially in the transboundary Volta River basin where different environmental data are collected and managed by different agencies in different countries and also where data access and dissemination are very challenging and difficult tasks, comprehensive river basin information systems are required. This paper presents the Oti River Basin Information System (OtiRBIS), a web-based data storage, management and analysis platform that addresses these needs and facilitates ILWRM implementation in the Kara river basin.*

## 1. INTRODUCTION

The integrated land and water resources management (ILWRM) is widely accepted as an appropriate approach to strive for sustainable water resources management and to adapt to impacts resulting from changing land use and climate [1]. To be efficient, the ILWRM applies a holistic system analysis of the river basin requiring appropriate knowledge including reliable data and information from different disciplines but also consistent tools for the storage, management and dissemination of relevant data and information [2, 3]. However, in most of river basins especially in West Africa, hydro-meteorological time series, information, and synthesized knowledge on water resources are diverse, often of variable quality, inconsistent resolutions, fragmented in nature and in many cases also inaccessible to a broad group of users impeding therefore sustainable water resources management [4, 5]. These challenges become much more remarkable and complex when dealing with meso-scale to macro-scale and transboundary basins. To overcome such challenges, sophisticated and user-friendly data management tools are needed.

Environmental Information Systems (EIS), are understood as an organized set of resources (staff, data, procedures, hardware, software,...) for collecting, storing, processing data and for delivering information, knowledge, and digital products [6, 7]. In the context of water resources management, these information systems are sometimes also called hydrological information systems (HIS).

A hydrological information system (HIS) can be defined as an integrated set of components that support the collection, processing, storage and dissemination of time series data on hydro-meteorological, hydrological, geo-hydrological and related variables with utmost efficiency along with safety and security features. The primary role of a HIS is to provide reliable data sets and information for planning, design and management of water resource and for research activities [6]. The main functions of a HIS are the integration of data from different sources and their archiving and management, the datasets documentation access via a catalogue, the visualization and download of data, the analysis of data (statistics, information crossing, simple GIS analyses) etc. [6]. According to [8], a HIS constitutes an important tool for sustainable water management.

HIS have been widely used to implement hydrological data management and showed the potential to be of high relevance in ensuring efficient management and sustainable use of water resources [9, 2, 10, 11, 12,13] but also in water related disasters reduction [14]. A web-based HIS can further act as a strategic gateway where scientists, citizens, stakeholders, and end users can securely use applications, stored information and services [10, 15] via web interfaces. For instance, AQUASTAT which is the FAO's global water information system developed by Land and Water Division (http://www.fao.org/nr/water/aquastat/main/index.stm), collects and disseminates information on water resources, water used and agricultural water management particularly in countries in Africa, Asia, Latin America and the Caribbean via a web interface. BASHYT (http://swat.crs4.it/) is a web-based GIS oriented information and support tool providing a set of application for data management, analysis and visualization for water resources management in the Black Sea Catchment [16]. Another example is the Consortium of Universities for the Advancement of Hydrologic Science, Inc (CUAHSI) Hydrologic information System (HIS) which is an internet-based system providing tools, standards and procedures for access to data for hydrologic analysis in the U.S.A [17].

In West Africa, few hydrological information systems are available. However, the existing ones lack functionalities needed for analysis and pre-processing of time series typically important for the integrated river basin analysis. For instance, a geoportal, the Volta Basin Authority (VBA) geoportal (http://131.220.109.2/geonetwork/srv/en/main.home) has been developed in order to improve access and integrated use of different information, promote multidisciplinary research and improve decision making for the Volta Basin. However, the geoportal lacks functionalities regarding time series visualization, analysis and management. Moreover, data on the VBA geoportal are not organized in a structured way which here refers to the organization of data according to each sub-catchment. However, as underlined by [18], the basin is the basic planning and management unit and appropriate ILWRM should take place at the basin scale, whether at the local catchment or aquifer, or at the transboundary river basin.

Located in West Africa in the Volta Basin's sub-catchment of the Oti Basin, the Kara River Basin (KRB) is a transboundary basin located between longitudes 0° 30' and 1° 38'E and latitudes 9°15'and 10° 01'N, with an area of 5,287 km$^2$ covering parts of Togo and Benin. Water resources in the basin are subject to increasing pressures from changing land use and climate, land cover and land degradation, and erosion. Currently, there is still a lack of adequate information to support effective river basin assessment and management. In fact, there are no proper catalogues on the available historical hydro-meteorological time series while some data are still in hard copy formats (paper records). In case data are digitally available, their formats differ from one agency to another. Moreover, the few existing data on soil, geology, land cover, socio-economic features on the basin are scattered. Complex data acquisition procedures make difficult any attempt to assess the KRB water resources and to implement sustainable management activities. A HIS is therefore necessary to collect, standardize and store the existing data on the basin in order to facilitate their access, dissemination and decision-making. [19].

This paper presents the OTI River Basin Information System (Oti RBIS) implemented based on the River Basin Information System (RBIS) software package [10, 16] and describes its potential as a platform for sustainable water resources management in the KRB.

## 2. THE RIVER BASIN INFORMATION SYSTEM (RBIS) SOFTWARE PLATFORM

As a software core of a HIS, the web-based river basin information system (RBIS) is a modular structured software platform with full read/write access developed in the Department of Geoinformatics, Hydrology and Modelling at the University of Jena (Germany). RBIS focuses on the management of environmental data (e.g. time series data, geospatial data), metadata, and the provisioning of standard compliant data exchange interfaces and services [11]. The following sections will give a brief overview about the system layout, main modules, applications of RBIS, and the integration with other ILWRM tools.

### 2.1. System layout

RBIS is based upon freely available open source software, ensuring a cost-efficient deployment and operation. The common layout of RBIS follows a 3-tier architecture. On the server side, the system is implemented using a standard Linux web stack with Apache web server, PHP programming language, PostgreSQL database management system (http://www.postgresql.org) and PostGIS extension (http://www.postgis.org) for spatial data support [11]. Moreover, MapServer (http://mapserver.org) is used for map rendering and OpenLayers (http://openlayers.org) as web map client. The JavaScript Library jQuery (http://jquery.com), the extension jQuery UI (http://jqueryui.com), the CSS-framework Bootstrap (http://getbootstrap.com) and some additional JavaScript Libraries are used for the front end. XML-files are used as description layers to generate different forms to view, search, edit and interlink stored datasets (for more detail see [16, 20]).

The main modules of RBIS are summarized in Figure 1.

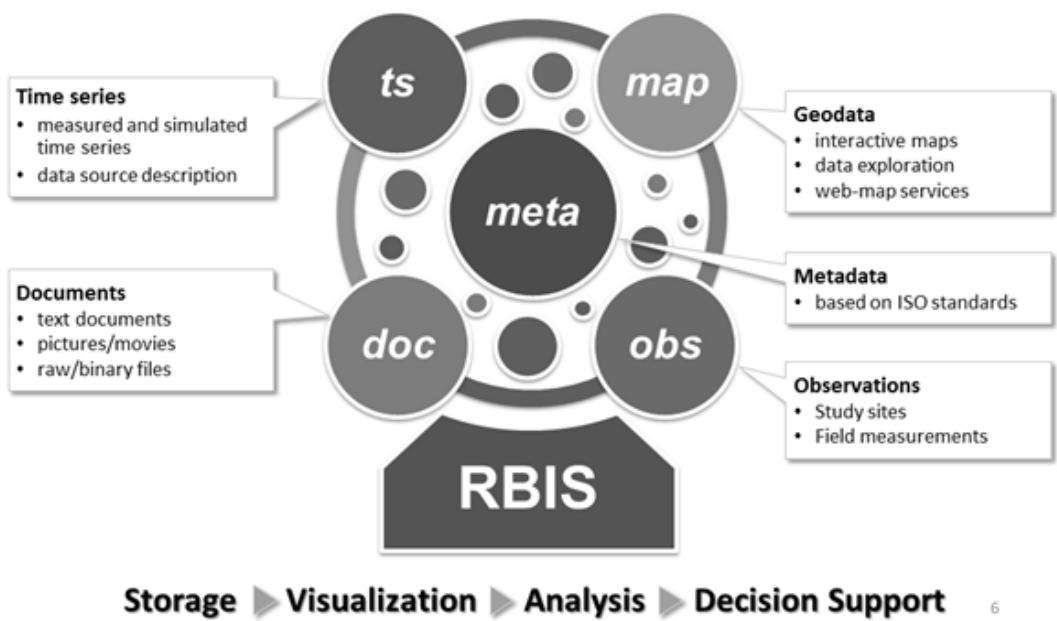

Figure 1: RBIS primary components [16]

In order to prevent unauthorized data manipulation or download access, RBIS offers a very fine grained user and permission management as one of its administrative core functionalities.

### 2.2. Application

RBIS focuses upon serving not only as data repository, data sharing platform and information system for environmental research projects of different sizes, but rather as an integral part of scientific workflows as it is e.g. needed for ILWRM. Therefore, RBIS provides several interfaces and services to access and discover data and metadata. For example an OGC standard-compliant CSW-Service (Catalogue Service for the Web) [21] has been set up to expose a catalogue of selected metadata records. Further, RBIS is an integrated part of the *Integrated Land Management System* (ILMS), a modular software platform that covers different steps of environmental systems analysis and planning in a flexible and user-friendly workflow [22]. In addition to RBIS, ILMS includes further tools such as (i) a software for the identification and classification of real-world objects from satellite imagery using methods of object based image analysis (ILMSimage), (ii) a software for the derivation of modelling entities based on GRASS GIS and QGIS (ILMSgis), and (iii) an environmental modeling framework for building, running and analyzing environmental simulation models, e.g. hydrological models (ILMSmodel).

RBIS has been applied for several environmental research projects (e.g. multi-disciplinary research projects or PhD projects). Two of them are also located in Africa, but in the southern part. The first one is the Okavango Basin Information System (OBIS) as a data and information management system for the Okavango Basin [16]. The second one is the SASSCAL RBIS (research project SASSCAL - Southern African Science Service Centre for Climate Change and Adaptive Land Management (http://sasscal.org)) containing information of several river basins in South Africa, Angola, Zambia, Namibia and Botswana.

## 3. THE IMPLEMENTATION OF THE OTI RBIS FOR ILWRM IN THE KARA RIVER BASIN

### 3.1. The Oti RBIS: an efficient tool for the storage, visualization and management of environmental data

The integrated river basin assessment as underlined above depends on the availability of reliable data and information on the basin. To meet this demand the Oti RBIS (figure 2) has been created. Though the initial RBIS is related to the Oti basin in general, the following section will present its importance for the Kara river basin, a sub-catchment of Oti basin in which ILWRM is currently being undertaken and has progressed more than tin other sub-catchments of the Oti basin.

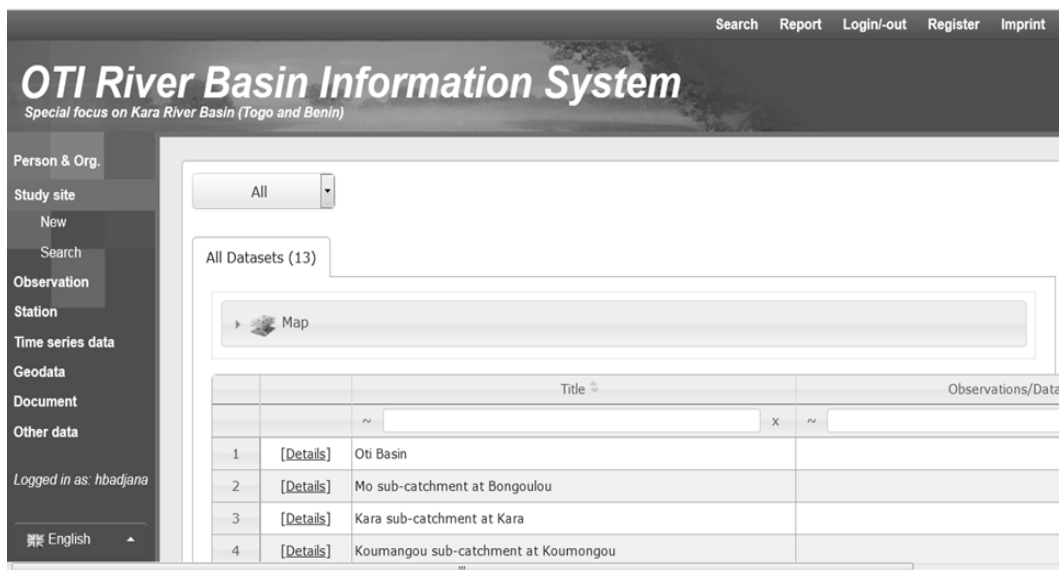

Figure 2: Screenshot of the Oti RBIS interface (study site overview)

The establishment of the Oti RBIS has allowed harmonizing and storing 112 hydro-meteorological time series that are available on the basin and collected from different sources, in a database. These time series are associated to measurement stations (e.g. gauging stations, climate and rainfall stations) which offer many other detailed information such as the elevation, the identification number of the station, the year of establishment, the responsible person or organization, the river or sub-catchment system to which the station belongs, or the year of establishment. Times series data once stored can be explored and displayed in order to identify the type of hydro-meteorological data available for each station and the spatial distribution of the gauging stations within the basin (figure 3). By means of the Oti RBIS, time series have been analysed for quality control. Artificial outliers have been detected and removed. All the data can be exported to standardized formats which has allowed to perform a climate trend analysis in the KRB, but also distributed hydrological modelling for climate and land use changes analysis.

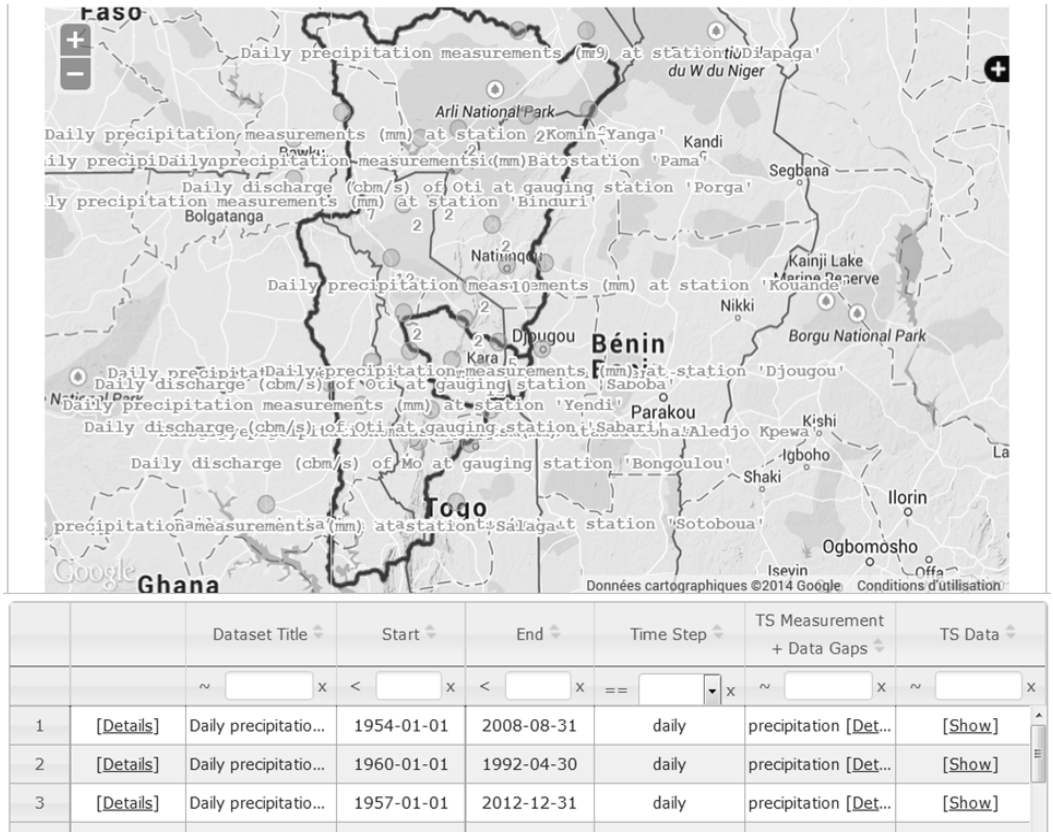

Figure 3: Screenshot of map and list view of stored measurement stations within the Oti RBIS

In addition, time series can be visualized individually (figure 4) and simultaneously for the length of times series and amount of missing data comparison between stations (figure 5). This helps to choose an overlap period for all the stations for a meaningful analysis. Basic indicators such as sum, maximum, mean, minimum or the trend can be calculated on demand for stored time series data at stations in and around a specified area (e.g. catchment or sub-catchment). Correlations analysis between stored times series of stations is another useful function that can be quickly applied in RBIS.

All the current existing time series stored in the Oti RBIS are at daily time step. Additionally to basics statistics, time series stored at daily step can easily be aggregated to monthly, yearly or hydrological year step with filled or unfilled gaps. Different text format and missing value codifications can be set while exporting, giving the possibility to easily process the data by external tools, e.g. by JAMS (Jena Adaptable Modelling System), a modular platform composed of different hydrological models for ILWRM [22] and currently in application in the KRB [23].

Besides the storage and management of time series, RBIS offers similar functions for geodata containing vector (ESRI Shapefile) and raster (GeoTIFF and JPEG) information and the metadata. Using this functionality, land use and land cover maps from the basin, catchments and sub-catchments boundaries, and their river networks shapefiles are stored in the Oti RBIS. These map and station layers are automatically linked so that desired time series within a specific catchments or sub-catchments can be visualized and exported. In addition, RBIS allows the storage and management of documents. All the results of socio-economic surveys within the basin have been processed and stored for subsequent use. The socio-economic surveys concerned available water sources, water-related ecosystem services and functions, safe water availability

20

and accessibility by local populations and their willingness to pay in case water becomes rare. The report on these surveys was uploaded to the Oti RBIS for the subsequent integrated assessment and decision making on ILWRM within the basin.

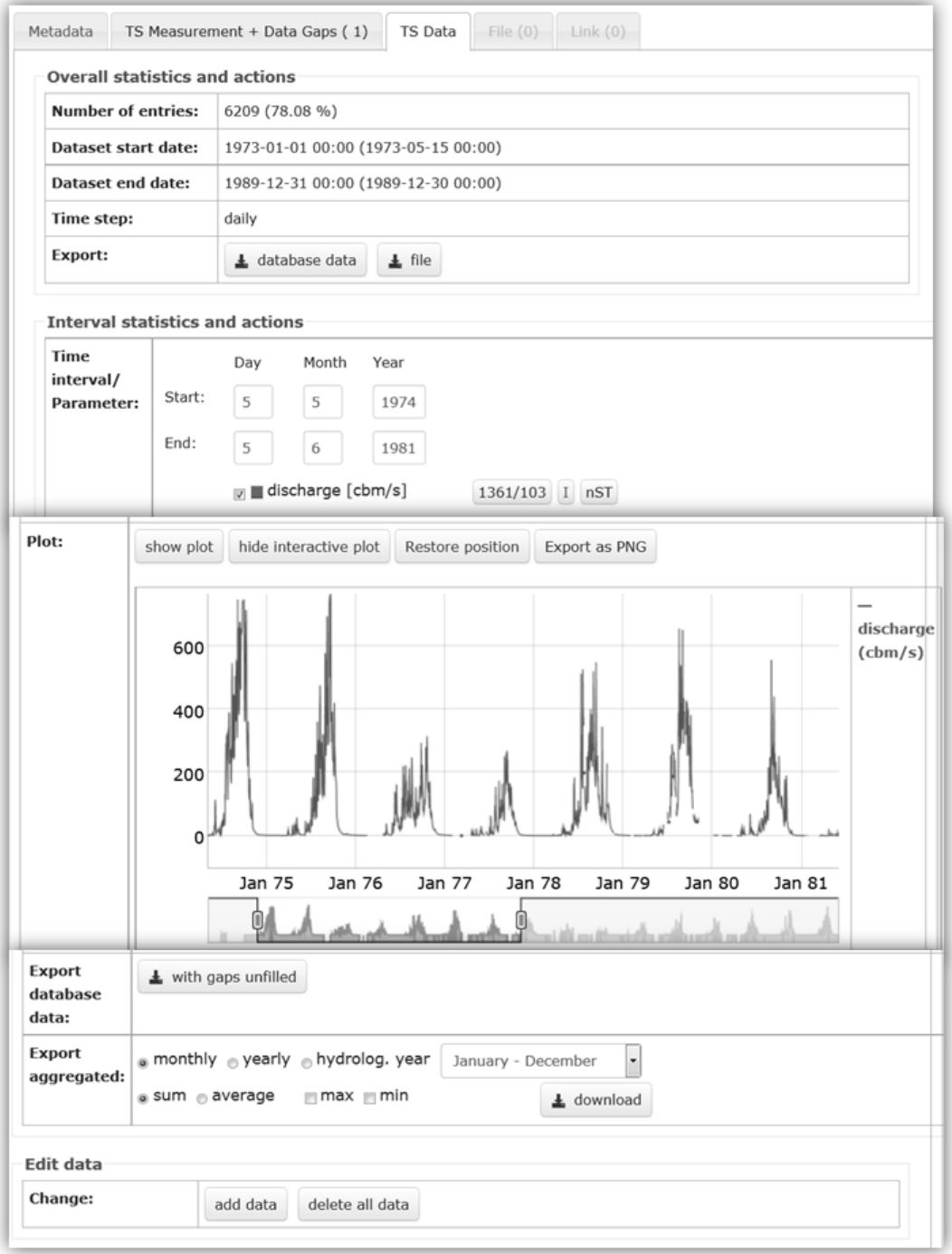

Figure 4: Screenshot of time series data detailed view showing simple statistics, possible actions and the interactive plot of discharge data measured at N'Naboupi station

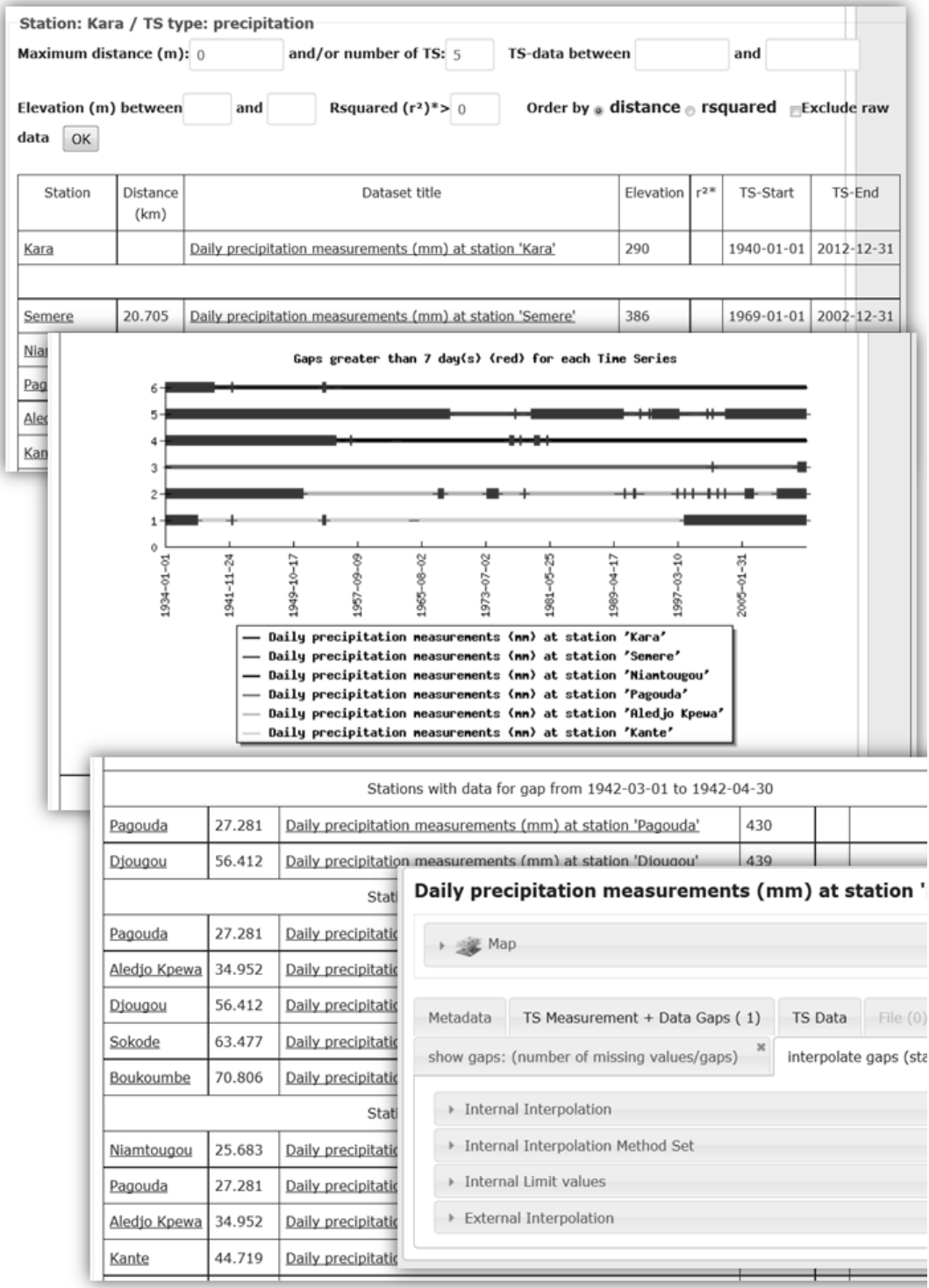

Figure 5: Availability of rainfall data and gaps (red parts) comparison between six stations of the Kara river basin.

## 3.2. RBIS for managing data quality

As a general problem in West Africa [24], the Kara River basin suffers from inadequate data due to the absence or incompleteness (in terms of length of records) of data or irregular observation periods, leading to a number of gaps or discontinuities in the time series.

The collected time series are stored together with detailed information about data gaps. RBIS provides multiple interpolation methods to fill these gaps. Available methods are composed of internal regression methods based on data from surrounding stations and external methods (figure 6). The complete time series thus constituted can be saved as corrected data in RBIS for subsequent use together with the used methods.

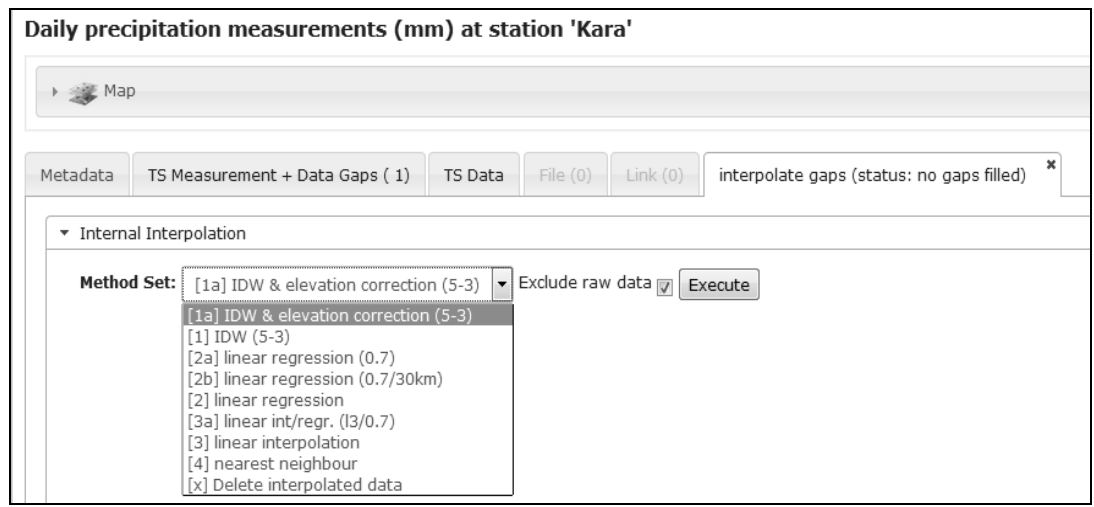

Figure 6: Available internal interpolation methods in RBIS

## 4. IMPLICATIONS

The availability of these functions in RBIS helped to easily identify to which extent the basin is currently gauged and what data are available for the basin. In fact only two synoptic stations (Kara and Niamtougou) within the basin and one rainfall station (Pagouda) outside but close to the boundary are still working while other stations in a limited number inside and around the basin are out of order. For the two synoptic stations that are still working only the Kara station provides data on evaporation and there is no available data on evaporation at that station since 1997. As far as the hydrological data are concerned, only flow data between 1954 and 1989 are available. As a general problem in Togo, most of river basins lack of river discharge data from 1990 till 2000s. This can be explained by the socio-political troubles in 1990s and the end of the ORSTOM (Office de la Recherche Scientifique et Technique Outre-Mer) mission in Togo at the end of 1980s. ORSTOM was a French research program which has allowed the installation of many synoptic and rainfall stations but also the equipments in many rivers throughout the country for the measurements of flows discharge and rate.

An integrated analysis of the data and information provided by the Oti RBIS indicates that the basin is poorly gauged and needs a rehabilitation of all measuring equipments in order to record reliable data for a proper integrated river system assessment that will support the sustainable management of water resources. The Oti RBIS becomes therefore a good tool for decision-making about the spatial distribution of the measuring equipments that will record the hydro-meteorological variables in the way that the collected data be representative of the basin climate and hydrology.

In addition, the collection, compilation and storage of available hydro-meteorological time series from multiple, distributed, autonomous, and heterogeneous data sources, countries and institutions, support the integrated access to and dissemination of data, an important step in the ILWRM. It is an important step made towards the solution to the challenges of data acquisition policies that differ a lot from one country to another and within a country from hydrological institutions to the meteorological ones. Moreover, in such a basin which is poorly investigated in terms of ILWRM, the time needed for data collection from different institutions, its transfer to digital formats and standardization constitutes a great challenge for any hydrological assessment. With the availability of Oti RBIS, this challenge is now addressed. Furthermore ILWRM as defined above requires the understanding of different relationships within the river basin system, which cannot be restricted to understanding interconnections within environmental systems but must go further to integrate scientific knowledge with information on social and economic systems [25]. The availability and possibility to integrate different data (climate, hydrological, geology, soils, land use, socio-economic, etc) in Oti RBIS constitute a particular advantage for undertaking a holistic or multidisciplinary approach to water resources management

The establishment of the Oti RBIS has been initiated within WASCAL project (West African Science Service Center on Climate Change and Adapted Land Use, www.wascal.org), a vast program funded by the German Federal Ministry of Education and Research (BMBF) that aims at building and merging adequate scientific and professional capacities in ten West African countries (Togo, Benin, Burkina-Faso, Ghana, Senegal, Mali, Niger, Nigeria, Côte d'Ivoire, Gambia) for the adaptation to changes in land management and climate. Since the target countries still face barriers such as the lack of data and information sharing systems, the transboundary issues (most of river basins are shared by two or more countries), the lack of cooperation and watershed-based integrated assessment and planning that poses significant challenges to the implementation of ILWRM. An information system such us the Oti RBIS is a suitable and useful tool to address these challenges. In fact the Oti RBIS once it is sufficiently loaded will constitute the first available and electronic catalogue of all hydro-meteorological and other water-related data in the basin. In addition, it will serve as a centralized system that provides sufficient data collected in different institutions and countries covered by the basin (Togo, Benin, Ghana, Burkina-Faso), providing information for decision making regarding ILWRM implementation at basin to sub-basin scales. This will also be a good mean to promote data and information exchange and trigger cooperation between different institutions and countries, a required process for effective ILWRM implementation.

## 5. CONCLUSION AND OUTLOOK

River basin information systems or in general hydrological information systems (HIS) are useful and necessary tools for ILWRM implementation efforts. Especially in West Africa where the availability of data and information for integrated hydrological systems' assessment is still challenging, accessible and integrated information systems will support the efforts towards sustainable management and development of water resources.

The establishment of the Oti RBIS is a great step towards any hydrological assessment in the basin whether at the sub-catchments scales or in the general Oti basin itself which remains very little explored in terms of integrated hydrological assessment. Hydro-meteorological time series and other water related data from multiple sources and different disciplines have been collected and processed; some data in raw format (hard copies) has been digitalized, compiled and stored after the standardization of all formats. This facilitates the access and dissemination of data as an important step in the ILWRM, offering the chance to promote collaboration between different agencies that collect hydro-meteorological data but also to further integrate watershed-based approaches in water resources assessment, management and planning.

Due to the underlying modular structure of the Oti RBIS it is very easy to extend it with regard to upcoming needs. The concept of study areas will also easily allow to extend the current data collection to other sub-catchments of the Oti River and potentially even to the Volta basins. Since the RBIS is currently under development in order to constitute a Decision Support System (DSS), such tools can also be considered for the integration in the Oti RBIS in order to provide DSS functionalities for the Oti basin, but also for other sub-catchments of the Volta basin.

**ACKNOWLEDGEMENTS**

This work has been funded by the German Federal Ministry of Education and Research (BMBF) through the West African Science Service Center on Climate Change and Adapted Land Use (WASCAL). Thanks to the Department of Geoinformatics, Hydrology and Modeling of the Friedrich-Schiller University of Jena (Germany) for developing RBIS.

**AUTHORS**

**Hèou Maléki Badjana** is currently a PhD student in the Graduate Research Program (GRP) of "Climate Change and Water Resources" of WASCAL (West African Science Service Center on Climate Change and Adapted land use) program. He received his Diploma (M.Sc.) in Environment Management at the University of Lomé, Togo. He is doing his PhD at the University of Abomey-Calavi (Benin) in collaboration with the Department of Geoinformatics, Hydrology and Modeling of the Friedrich-Schiller University of Jena, Germany. His research interests are river basin assessment and modeling, climate analysis, integrated land and water resources management.

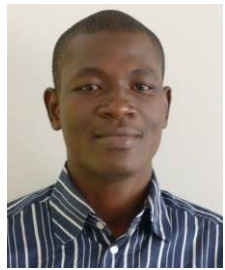

**Franziska Zander** received her diploma degree in Geography (2007) and is Research Assistant at the Department of Geoinformatics, Hydrology and Modeling of the Friedrich-Schiller University of Jena, Germany. She is involved in the development and administration of the environmental data information system RBIS and works for several international and multidisciplinary research projects to support research data management (e.g., projects in Vietnam (LUCCi), Africa (TFO, SASSCAL), Brazil or Chile). Her research interests are time series data and geodata management, research data management and web-mapping with Open Source-GIS-Tools. 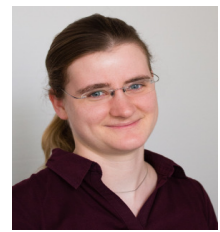

**Sven Kralisch** is a Senior Researcher at the Department of Geography / Geographic Information Science of the University of Jena, Germany. After studying Computer Science he started working in the area of environmental modelling and integrated water resources management with a strong focus on methodological and technical aspects. After receiving his PhD in 2004, Dr. Kralisch participated in numerous national and international research projects in Europe, southern Africa and Brazil. As an expert for environmental simulation frameworks, data management systems and service-oriented software architectures, his main research interest is on the design and implementation of integrated, Open-Source software systems that support scientists, planners and decision makers in assessing the impact of land management and climate change on environmental systems. 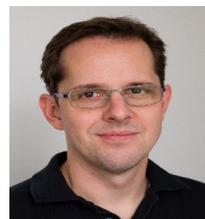

**Jörg Helmschrot** holds a diploma in Geography (1999) and a PhD in Geography and Geoinformatics (2006), both from the Friedrich-Schiller-University in Jena. In 2010/11 he did his Postdoc at Mountain Hydrology Research Group at the University of Washington. In his current position as senior researcher and scientific coordinator of SASSCAL at University of Hamburg, Dr. Helmschrot is involved in the coordination of water and climate research, the implementation of a regional weather monitoring network, the establishment of the SASSCAL Open Access Data Center and hydrological research in SASSCAL for decision support. As an expert for integrated analysis, modelling and assessment of hydrological systems at different scales (hill slope, catchment, ecosystem, landscape scale) and process-oriented modelling of catchment and wetland systems as well as their assessments, he has been involved in numerous national and international research projects with different aspects of hydrological research and integrated land and water resources management (ILWRM) in semi-arid regions, in particular in Southern Africa, but also in Turkey, the USA, Tibet and Australia. 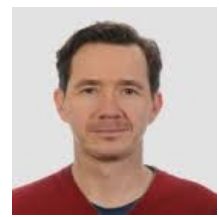

**Wolfgang-Albert Flügel** is a retired university professor and former head of the Department for Geoinformatics, Hydrology and Modelling (DGHM) at the Friedrich-Schiller University of Jena, Germany. He is an international recognized scientist on the field of Geoinformatics and Hydrology with numerous contacts to national and international research teams and institutions and more than 90 scientific publications as single author and in co-authorship. From 1985 until 1990, Prof. Flügel was working as Senior Specialist Scientist at the Hydrological Research Institute (HRI) in South Africa and from 2002 until 2003 he was Principal Hydrologist at the International Water Management Institute (IWMI) in Colombo, Sri Lanka. During his academic career he was visiting professor at universities in Mexico, South Africa, India and the US and organizing member of expeditions to the Canadian Arctic and the Antarctic. His main research interests are applied Geoinformatics in process hydrology and hydrological systems analyses, regionalization and GIS, hydrological basin modelling and the development of integrated Decision Information Support Tools (DIST). His research activities focus on: (i) Integrated Land and Water Resources Management (ILWRM); (ii) Distributed hydrological modelling; (iii) Regionalization and regional multi-scale analysis of climate change; (iv) Climate Change impact assessment and analysis for adaptive ILWRM strategies; (v) Sustainable irrigation management; (vi) Dryland and irrigation salinity research. In realizing his research interests, he had carried more than 90 international projects with national, European and international funding in Southern Africa, Antarctic, Bhutan, Canadian Arctic, Australia, Brazil, China, Europe, India, Nepal and Turkey. 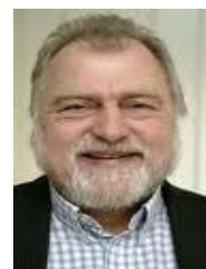